# CNN-Based Deep Architecture for Reinforced Concrete Delamination Segmentation Through Thermography

Chongsheng CHENG[1], Zhexiong SHANG[2] and Zhigang SHEN[3]

[1] The Durham School of Architectural Engineering and Construction, University of Nebraska-Lincoln, 113 Nebraska Hall, Lincoln, NE 68588-0500; e-mail: cheng.chongsheng@huskers.unl.edu

[2] The Durham School of Architectural Engineering and Construction, University of Nebraska-Lincoln, 113 Nebraska Hall, Lincoln, NE 68588-0500; e-mail: szx0112@huskers.unl.edu

[3] The Durham School of Architectural Engineering and Construction, University of Nebraska-Lincoln, 113 Nebraska Hall, Lincoln, NE 68588-0500; e-mail: shen@unl.edu

**ABSTRACT**

Delamination assessment of the bridge deck plays a vital role for bridge health monitoring. Thermography as one of the nondestructive technologies for delamination detection has the advantage of efficient data acquisition. But there are challenges on the interpretation of data for accurate delamination shape profiling. Due to the environmental variation and the irregular presence of delamination size and depth, conventional processing methods based on temperature contrast fall short in accurate segmentation of delamination. Inspired by the recent development of deep learning architecture for image segmentation, the Convolutional Neural Network (CNN) based framework was investigated for the applicability of delamination segmentation under variations in temperature contrast and shape diffusion. The models were developed based on Dense Convolutional Network (DenseNet) and trained on thermal images collected for mimicked delamination in concrete slabs with different depths under experimental setup. The results suggested satisfactory performance of accurate profiling the delamination shapes.

## 1. INTRODUCTION

Delamination as the horizontal crack embedded in the subsurface of the bridge deck is often a resultant of corrosion-induced deterioration of reinforcement. Its continuous progress will eventually affect the structural integrity of the bridge deck (Gucunski and Council 2013). Thus, profiling its extent is essential in the point of view of structural health monitoring (SHM). Practically, determination of its location, boundary (shape), and area (ratio over the entire deck) are typically desired. Compared to the conventional method such as hammer sounding and chain dragging, nondestructive detection (NDT) methods such as Infrared Thermography provided a fast and effective way to detect shallow delamination of bridge deck (Dabous et al. 2017; Kee et al. 2011; Maierhofer et al. 2006; Omar and Nehdi 2017; Washer et al. 2013). Studies revealed the mechanism of detection through thermography was based on the principle of the developed temperature contrast between delaminated and intact areas during a day. Thus, the focus of previous researches was on investigating the detectability based on temperature contrast variations in terms of different environmental conditions and configurations of delamination geometry (Hiasa et al. 2017; Hiasa et al. 2017; Sultan and Washer 2017). So far, the image processing methods proposed in the literature were relied on the optimal threshold selection (Sultan and Washer 2017), k-mean clustering (Omar and Nehdi 2017), and region regrowth based on contrast (Abdel-Qader et al. 2008; Ellenberg et al. 2016). Given the context of profiling the delamination boundary (shape), there were limited studies to utilize the spatial features of delamination through thermography.

In this paper, we introduce the Convolutional Neural Network (CNN) based deep learning architecture to tackle the problem. This framework has been reported that overperformed conventional methods in object detection and segmentation in computer vision which the objects in an image could be represented by features in the more abstract levels (LeCun et al. 2015). Inspired by densely connected CNN architecture (DenseNet) for image semantic segmentation (Huang et al. 2017), the delamination profiling (segmentation) in the thermal image could be addressed through pixel-wised labeling under supervised learning scheme. The purpose of this study aims to investigate the performance of this architecture in delamination segmentation under experimental settings.

## 2. RELATED WORK

Many CNN-based applications were found in the field of medical image processing for tissue segmentation (Pekala et al. 2018; Schlegl et al. 2017). It is rare to find such applications in concrete delamination NDT through thermography. This paper focused on the architecture proposed by Huang et al. (2017), which a dense block module was introduced. The block module consisted a sequence of densely connected convolutional layers that the information not only propagated through layer sequentially but also concatenated to the later layers. In this way, the features could be re-used, and the variations were increased so that the model were easy to train and highly parameter efficient. Between dense blocks, the transition layer was used which consisted of batch normalization and 1x1 convolutional layer then a max pooling layer. This architecture has been evaluated on several benchmark datasets (e.g. CIFAR-10, CIFAR-100, SVHN and ImageNet) and significant improvement was found over the state-of-the-art at the time for classification tasks. Although the original DenseNet was designated for classification, it could be extended for semantic segmentation with 'end-to-end' scheme. Based on its key module of the dense block, the architecture was then built for our delamination segmentation task.

## 3. METHOD
*Architecture Development*

Figure 1 shows the proposed architecture with three dense blocks followed by three upscale structures. Each dense block had four convolutional layers inside connected and concatenated so that the number of filters on output will quadruple from the input. In this way, the maximum information and gradient flow are preserved in each dense block and easy for training (Huang et al. 2017). The input firstly passed a convolutional layer and max pooled followed by batch normalization (BN) for low level feature extraction. Then three dense blocks were followed with three transitional layers in between to decrease the spatial size so that a feature map with a size of 20x20x512 was generated for embedded representation. Based on this scheme, the feature tends to be represented by the latent variables. Before upscaling, the intermediate layer and the dropout layer were used to against potential overfitting problem within convolution layers (Wu and Gu 2015). To propagate back to original size as input, three upscale blocks were used based on bilinear interpolation (Long et al. 2015) for up-sampling followed by a 3x3 convolutional layer so the "end-to-end" scheme could be achieved for pixel-wise segmentation.

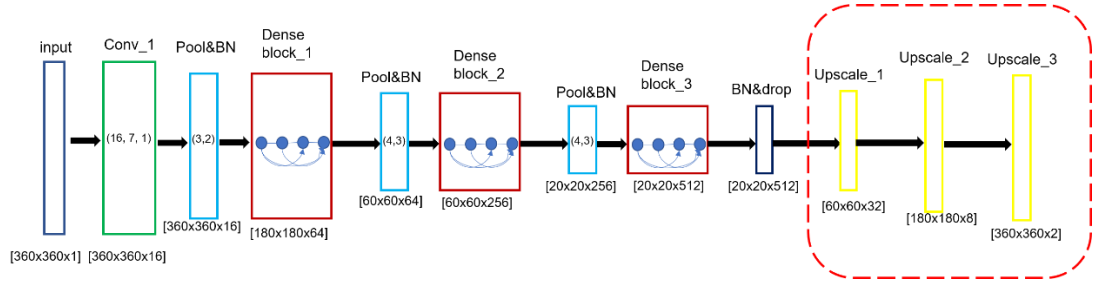

**Figure1. Developed architecture**

## 4. EXPERIMENTS
*Design and Setup*

Experiment samples were designed to simulate the delamination in concrete slab with different depths. Figure 2 (b) shows the layout of reinforcement and position of mimicked delamination. Three slabs had been constructed and heated outside from 10 am to 2 pm by the sun in three adjacent days (Figure 2c). The slab was then moved inside for heat releasing and data was recorded by the thermal camera at a sampling rate of 0.1Hz (Figure 2a). Four cases were conducted to train 4 models for supervised learning in four situations (Table 1). Training size (image height by width by frame) for each case was then determined by the criterion of contrast limiting and data augmentation which are discussed in the following sections.

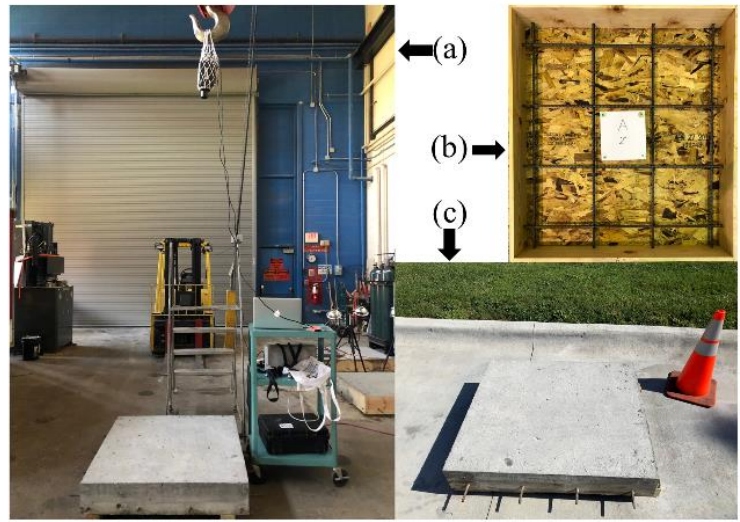

**Figure2. (a) Data collection; (b) slab with mimicked delamination; (c) slab heated at outdoor by the sun**

**Table 1**

| Case | Slab Name | Depth from top | Slab Size | Delamination Size | Original size | Training Size |
|---|---|---|---|---|---|---|
| 1 | A | 1.75" | 40'x45' | 10'x10' | 639x562x445 | 360x360x4005 |
| 2 | B | 2.75" | 40'x45' | 10'x10' | 660x578x413 | 360x360x3717 |
| 3 | C | 3.75" | 40'x45' | 10'x10' | 632x553x511 | 360x360x4599 |
| 4 | A, B, and C | all | 40'x45' | 10'x10' | ? x? x1369 | 360x360x12321 |

*Data Preparation and Description*

The inclusion of training data was determined by using the contrast limits. Here we calculated the temperature contrast between the delaminated and sound areas, and then the threshold of 0.5 ºC was chosen as the cut-off to determine the length of training data for each case (Figure 3). The selection of this threshold was based on (1) value was recommended by ASTM standard (ASTM 2007); (2) author's experience that the feature variation of mimicked delamination deviated too far away when it was smaller than the threshold. As the results, 445, 413, and 511 thermal images were included for case 1, 2, and 3; and case 4 included a summation of previous three cases.

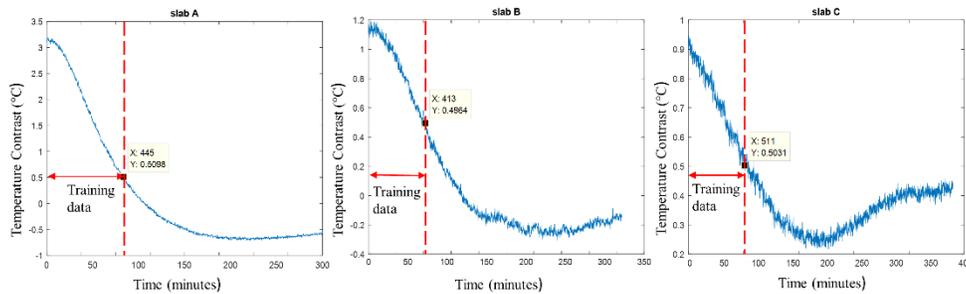

**Figure 3. Data selection of three slabs for training by contrast limit**

Figure 4 shows how the thermal feature of delamination varied for each slab during the heat releasing stage. At Figure 4 (a), the feature of delamination differs in contrast (temperature difference) in three slabs. The shallower depth of delamination located (slab A), the larger contrast could be observed. Also, a high temperature zone on edges was observed due to the closeness to the boundary of the slab. Figure 4 (b) shows the section profile of each slab at different time windows which the variation of shape could be observed. The solid red line in each subplot (Figure 4 b) indicated the truth boundary location for each delamination buried. Thus, the object of this study could be furtherly refined to train the model to capture these variations so that the true shape and size of delamination could be represented.

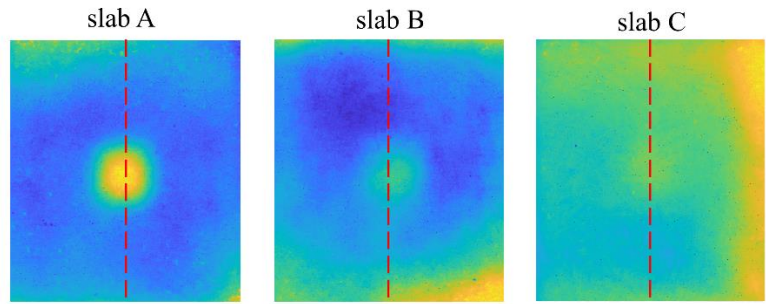

(a)

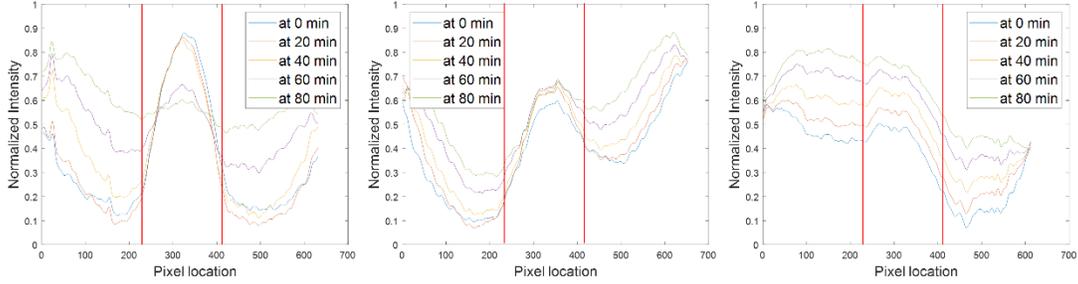

**Figure 4. Raw thermal image of three slabs at time 0 (a); and temperature variations at different time windows during heat releasing (b)**

*Data Augmentation and Labeling*

One of the issues with machine learning/deep learning models is over fitting when the training sample is limited in size and generality. To increase the training size and variety, data augmentation is an often used technology (Mikołajczyk and Grochowski 2018). Conventional augmentation is to use the combination of transformations which includes: crop, translation, rotation, reflection, scaling, and shearing. Here we focused on the crop and translation at the current stage of the study. Figure 5 shows the augmented data for a single image and its corresponding labels. After the augmentation, each raw thermal image had been divided into 9 cropped images with a size of 360x360 in pixels. This addressed the issue that the designed the delamination was always on center which may cause location sensitive of trained models. Then each image was labeled in a way that 1 represents the delaminated area and 0 as the background in the pixel level. Thus, the sample size of training data increased dramatically and was presented in the last column of table 1.

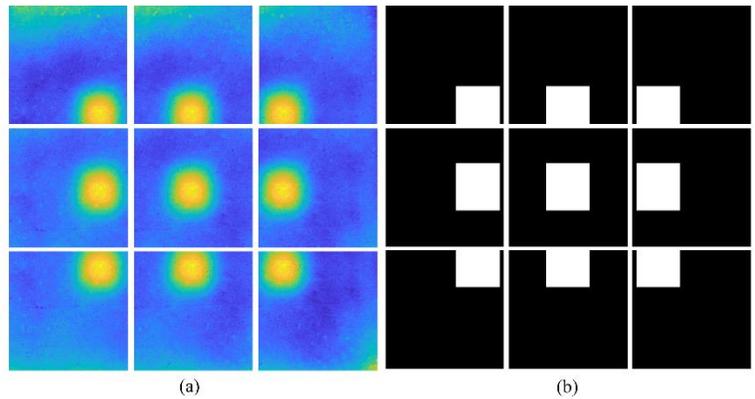

(a)          (b)

**Figure 5. (a) Data augmentation through crop and translation; (b) the corresponding labels**

*Training and Parameter Setting*

Four models were trained individually by using the training sets in four cases shown in table 1. For each case, the data was split into 70% for training and 30% for testing. Total iteration (epoch) of training was set to 40 and mini-batch was deployed with the size of 32 (Li et al. 2014). The activation function of ReLu was selected for all embedded CNN layers. The loss function is defined by softmax activation paired with cross-entropy penalty at the output layer for minimization. The optimization was then carried by Adam optimizer with a learning rate of 0.0001. Additionally, the L2 regularization was implemented to handle the overfitting potential by considering the weight of model complexity. All models were trained and tested by using Python with TensorFlow library on GPU.

5. **RESULTS**

To evaluate the performance of model training, two metrics were used: true positive rate (TP) and false negative rate (FN). TP was defined to measure the rate of true delamination over predicted delamination. FN calculated the rate of false background over the predicted background. We would expect as higher as possible in TP for "precision", and as lower as possible in FN as the "fall-out". Table 2 shows the results after 40 iterations of training. All cases returned TP rate over 99% and low FN rate which indicated a good performance of current models in terms of accuracy. Data visualization was combined with validation shown in Figure 6.

**Table 2**

| Case | Epoch | Train TP | Train FN | Test TP | Test FN |
|---|---|---|---|---|---|
| 1 | 40 | 0.9949 | 0.00006 | 0.9944 | 0.00011 |
| 2 | 40 | 0.9967 | 0.00004 | 0.9962 | 0.00006 |
| 3 | 40 | 0.9952 | 0.00006 | 0.9942 | 0.00009 |
| 4 | 40 | 0.9979 | 0.00003 | 0.9979 | 0.00004 |

*Validation with Down Scaled Images*

The data was visualized and validated by downscaling the original thermal images for all cases (shown in Figure 6). For all four cases, the performance degraded in the later time window which the pattern of delamination was largely diffused in raw images (comparing result between time 0 and time 67). Thus, the miss alarmed segmentation was more intended to occur at the later image in case 1, 2, and 3. Comparing case 1 to 3, the depth affected significantly for shape profiling even the architecture of the three cases was the same. The deeper delamination buried, more degradation in performance for shape sketching (e.g. case 3). Case 4 showed the best overall shape profiling performance due to the combination of training samples.

**6. DISCUSSION AND CONCLUSIONS**

Although all cases showed a good performance in terms of TP and FN rates, the shape profiling in validation still needs improvement. One issue could be the unbalanced samples due to the layout of slabs since the total area (pixels) for the background were much larger than the area (pixels) assigned to the delamination. Thus, there are more training samples for than delamination which makes the model has more experience on learning background. Also, the unbalanced classes might also cause model bias during training which will be considered in the future development. It is our finding that increasing the sample variety indeed improved the model generality (better shape profiling) for case 4. It may reveal the potential power of deployed architecture and deep learning technology. In the future work, the parameter tuning, architecture refining, more data augmentation will be conducted.

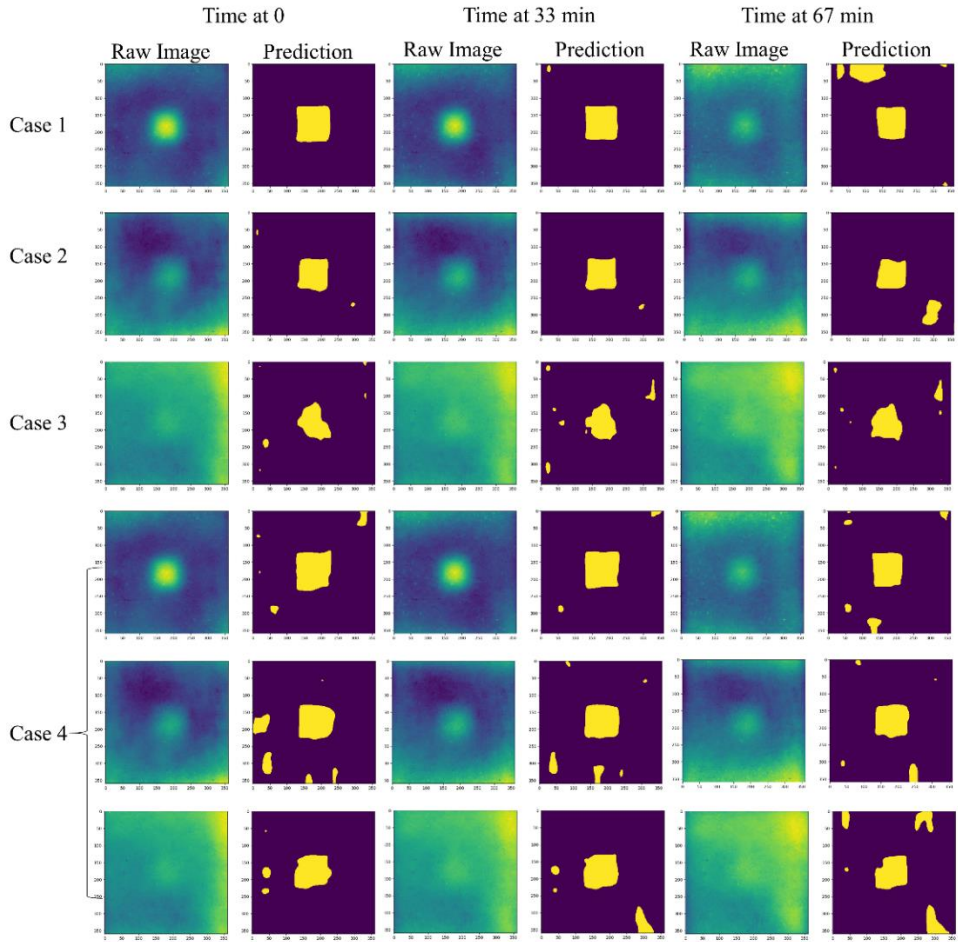

**Figure 6 Validation and visualization of results for four cases**